\documentclass[10pt, conference, letterpaper]{IEEEtran}
\IEEEoverridecommandlockouts
\usepackage{cite}
\usepackage{amsmath,amssymb,amsfonts}
\usepackage[noend]{algpseudocode}
\usepackage{subcaption}
\usepackage{algorithm}
\usepackage{graphicx}
\usepackage{textcomp}
\usepackage{booktabs}
\usepackage{xcolor}
\usepackage{comment}
\def\BibTeX{{\rm B\kern-.05em{\sc i\kern-.025em b}\kern-.08em
    T\kern-.1667em\lower.7ex\hbox{E}\kern-.125emX}}
\begin{document}

\title{Applying Graph-based Deep Learning To Realistic Network Scenarios}

\author{
\IEEEauthorblockN{Miquel Ferriol-Galmés\IEEEauthorrefmark{1},
José Suárez-Varela\IEEEauthorrefmark{1},
Pere Barlet-Ros\IEEEauthorrefmark{1} and
Albert Cabellos-Aparicio\IEEEauthorrefmark{1}}
\IEEEauthorblockA{\IEEEauthorrefmark{1}Barcelona Neural Networking Center, Universitat Politècnica de Catalunya\\Email: \{mferriol, jsuarezv, pbarlet, acabello\}@ac.upc.edu}
}

\maketitle

\begin{abstract}
Recent advances in Machine Learning (ML) have shown a great potential to build data-driven solutions for a plethora of network-related problems. In this context, building fast and accurate network models is essential to achieve functional optimization tools for networking. However, state-of-the-art ML-based techniques for network modelling are not able to provide accurate estimates of important performance metrics such as delay or jitter in realistic network scenarios with sophisticated queue scheduling configurations. This paper presents a new Graph-based deep learning model able to estimate accurately the per-path mean delay in networks. The proposed model can generalize successfully over topologies, routing configurations, queue scheduling policies and traffic matrices unseen during the training phase.
\end{abstract}

\section{Introduction}
Many typical network optimization problems are known to be NP-hard (e.g. load balancing \cite{hartert2015declarative}, QoS routing \cite{wang2011study}). Thus, we have witnessed the use of different traditional optimization algorithms to address such kind of problems (e.g., ILP, SGD). In essence, a network optimization tool can be achieved by combining two main elements: a network model, and an optimization algorithm. In this context, the accuracy of the model is critical to achieve high-quality results, as it is the one in charge of estimating the resulting performance after changing a configuration parameter in the network~\cite{rusek2019unveiling}.

Some network optimization problems (e.g., load balancing) can be solved using relatively simple network models (e.g., based on fluid models), while others (e.g., end-to-end delay optimization) require much more complex models, such as packet-level network simulators. However, these complex models are computationally very expensive and, as a result, do not meet the requirements to achieve online network optimization in large-scale network scenarios.

Alternatively, many analytic network models have been developed in the past~\cite{ciucu2012perspectives}~\cite{el2006optimal}. However, such models make strong assumptions that do not hold in real-world networks, for instance neglecting queuing delay or probabilistic routing. In addition, they are unable to accurately model scenarios involving arbitrary sequences of complex queuing policies \cite{xu2018experience}. As a result, they are not accurate for large networks with realistic routing and queuing configurations and, consequently, they are not practical for modelling relevant performance metrics like delay, jitter or loss in real-world networks.

This issue has attracted the interest of the networking community, which is recently investigating the application of Deep Learning (DL) techniques to build efficient networks models, particularly focused on complex network scenarios and/or performance metrics. Researchers are using neural networks to model computer networks \cite{guo2014survey} and using such models for network optimization, in some cases in combination with advanced strategies based on Deep-Reinforcement Learning \cite{huang2019self,chen2019deeprmsa}.

Most of existing DL-based solutions for network modelling, mainly rely on common Neural Network (NN) architectures such as feed-forward NNs, or Convolutional NNs. However, data from computer networks is essentially represented in a graph-structured manner, and this kind of NNs are not suited to learn data structured as graphs. As a result, they have very limited applications. For instance, they cannot be trained in a set of networks and then operate successfully in other networks, nor understand routing and queueing policies different from those seen during the training phase. All this is the result of their poor generalization capability over graphs.

In this context, Graph Neural Networks (GNN)~\cite{scarselli2008graph} have recently emerged as effective techniques to model graph-structured data. Particularly, these new types of neural networks are tailored to understand the complex relationships between connected elements in graphs. More in detail, the internal architecture of a GNN is dynamically built based on the elements and connections of input graphs, and this permits to learn generic modelling functions that do not depend on the graph structure. As a result, GNN has demonstrated an outstanding capability to generalize over graphs of variable size and structure in many different problems \cite{gilmer2017neural, battaglia2016interaction, zhou2018graph}.

In this paper we present a new GNN model that is able to predict the per-path delay given an input topology, routing configuration, queue scheduling policy, and traffic matrix. This architecture is easily extensible to estimate other relevant performance metrics such as jitter or packet loss. More importantly, it offers a strong relational inductive bias over graphs~\cite{battaglia2018relational}, which endows the model with strong generalization capabilities over topologies, routings, queuing policies and traffic unseen during the training phase.

\section{Graph-based Deep Learning network model}
In this section, we present a GNN-based model that introduces queue scheduling as an inherent component of the neural network architecture. This enables to model accurately how different queueing configurations (scheduling and queue parameters) affect the network performance. The model architecture is designed to generalize to different networks never seen during the training phase. We refer the reader to \cite{scarselli2008graph, gilmer2017neural, battaglia2018relational} for a comprehensive background on GNN.

\subsection{Overview}
The main intuition behind this architecture is as follows. The model considers three main entity types: $(i)$ the queues present at output ports of network devices, which have some given configuration parameters (size, priority, weight, etc.). In total, each node has on each port between 2-5 queues (or 1 queue if it implements FIFO), $(ii)$ the links from the topology (i.e., connections between nodes), which include their capacity as input features, and $(iii)$ the paths formed by the input routing scheme. Thus, the per-source-destination traffic of the input traffic matrix is encoded in the initial path states. Particularly, they contain information about the average bandwidth generated on that path.

\begin{figure}[!t]
\centering
\includegraphics[width=0.8\columnwidth]{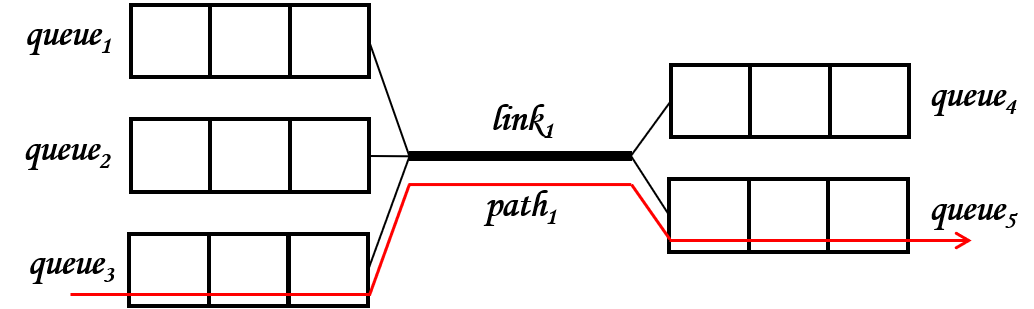}
\caption{Schematic representation of the our network model.}
\label{fig:architecture_overview}
\end{figure}

Figure~\ref{fig:architecture_overview} shows a scheme that represents how the GNN treats these three components. First, the state of paths depend on the concatenation of queues and links they traverse. In this case, $path_1$ follows the sequence: $[queue_3, link_1, queue_5, ...]$. %Understanding paths as the sequence of queues and links is a fundamental key to the generalizability of the model.
At the same time, the state of queues and links depend on all the paths that cross them. Hence, there is a circular dependency between the states of these three elements that the GNN model should resolve to eventually produce delay estimates on each path.

%The next sections provide a mathematical description of the novel GNN architecture that introduces queue scheduling as an inherent component of the neural network architecture. This new architecture enables to model accurately how the relevant network parameters affect the performance.

\subsection{Notation}
%- Follow Krzysztof notation...
A computer network can be defined by a set of queues $Q = {q_i}, i \in (0,1,...,n_q)$ and the links that connect them $L = {l_j}, j \in (0,1,...,n_l)$. Let us also consider the function $j(l_j)$ that returns all the queues that inject traffic into link $j$. The routing scheme in the network can be described by a set of paths $P = {p_k}, k \in (0,1,...,n_p)$. Each path traverses a sequence of queues and links $p_k = (q_{k(0)},l_{k(0)},...,q_{k(|p_k|)},l_{k(|p_k|)}$, where $k(i)$ is the index of the i-th link or queue in the path $k$. The properties (features) of queues, links and paths are denoted by $x_{q_i}$, $x_{l_j}$ and $x_{p_k}$ respectively. In this particular case, the initial features of queues ($x_q$) are the size, the priority and the weight. The initial link features ($x_l$) are the capacity and the scheduling policy of the node. Finally, the initial path features ($x_p$) are defined by the bandwidth generated for each source-destination pair.

\subsection{Network Model}
%- Add algorithm of the GNN model
%- Nota: No hablar nunca de extensión de RouteNet. Es un "novel GNN model"
%Let us consider the delay on path $p_k = {(q_{k(0)},l_{k(0)},q_{k(1)},l_{k(1)},...)}$.  In a secenario where the packet loss is negligible, the delay of each path could be computed as $\sum_{i} d(q_{k(i)})+ d(l_{k(i)})$ where $d(q_m)$  represents the delay on the m-th queue and $d(l_m)$ the delay on the m-th link found in the path.  However, a scenario where packets losses are negligible can not be considered as a real scenario. This, introduces a sequential dependece between the queues and links states.

We describe the state of the queues, links and paths as $h_q$, $h_l$ and $h_p$ respectively. These unknown hidden vectors that describe the state are expected to contain some meaningful information about links (e.g., utilization), queues (e.g., load, packet loss rate), and paths (e.g., end-to-end metrics such as delays or packet losses). Considering the aforementioned assumptions, the following principles can be stated:
\begin{enumerate}
  \item The state of a path depends on the states of all the queues and links that it traverses.
  \item The state of a link depends on the states of all the queues that inject traffic in this link.
  \item The state of a queue depends on the states of all the paths that inject traffic in it.
\end{enumerate}

These principles can be formulated by the following equations:
\begin{equation}
\label{eq:f_q}
h_{q_i} = f_q(h_{p_1},...,h_{p_m}), q_i \in p_k, k = 1,...,j
\end{equation}
\begin{equation}
\label{eq:f_l}
h_{l_j} = f_l(h_{q_1},...,h_{q_m}), q_m \in j(l_j)
\end{equation}
\begin{equation}
\label{eq:f_p}
h_{p_k} = f_p(h_{q_{k(0)}},h_{l_{k(0)}},...,h_{q_{k(|p_k|)}},h_{l_{k(|p_k|)}})
\end{equation}
where $f_q$, $f_l$ and $f_p$ are some unknown functions.

A direct approximation of functions $f_q$, $f_l$ and $f_p$ is complex given that: $(i)$ Equations \ref{eq:f_q}, \ref{eq:f_l} and \ref{eq:f_p} define a complex nonlinear system of equations with the states being hidden variables, $(ii)$ these functions depend on the input routing scheme, the mapping of traffic flows to queues (Type of Service) and the different queue policies in the network, and $(iii)$ the dimensionality of all the possible states is extremely large.

GNNs have shown an outstanding capability to work as universal approximators over graphs. With this, the proposed GNN architecture finds an approximation for the $f_q$, $f_l$ and $f_p$ functions that can be applied to unseen topologies, routing schemes and queue policies.% but still understands it. For this, the proposed architecture is based on message-passing neural networks (MPNN) \cite{gilmer2017neural} which is a specific GNN family with previous applications in the networking field \cite{rusek2019unveiling} \cite{suarez2019challenging} \cite{geyer2017performance}.

\subsection{Proposed GNN Architecture}

Algorithm \ref{alg:architecure} describes the internal architecture of the GNN. This architecture takes advantage of the ability of GNNs to meet the challenges presented. Particularly, it solves the circular dependencies described in Equations \ref{eq:f_q}, \ref{eq:f_l} and \ref{eq:f_p} by executing an iterative message passing process. In each message passing iteration, graph elements exchange their hidden states $h_q$, $h_l$ and $h_p$ with their neighbours according to the operations in Algorithm~\ref{alg:architecure}, and this is repeated $T$ iterations (loop from line 4). Thus, hidden states $h_q$, $h_l$ and $h_p$ eventually should converge to some fixed points.

\algnewcommand\algorithmicforeach{\textbf{for each}}
\algdef{S}[FOR]{ForEach}[1]{\algorithmicforeach\ #1\ \algorithmicdo}
\begin{algorithm}[!t]
\caption{Internal architecture of the proposed GNN model}
\hspace*{\algorithmicindent} \textbf{Input:} $x_q$, $x_l$, $x_p$, $P$ \\
\hspace*{\algorithmicindent} \textbf{Output:} $h^T_q$, $h^T_l$, $h^T_p$, $\hat{y_p}$
\begin{algorithmic}[1]
\ForEach {$q \in Q$} $h^0_q \gets [x_q,0...0]$ \EndFor
\ForEach {$l \in L$} $h^0_l \gets [x_l,0...0]$ \EndFor
\ForEach {$p \in R$} $h^0_p \gets [x_p,0...0]$ \EndFor
\For{t = 1 to T}
    \ForEach {$p \in P$}
        \ForEach {$q,l \in p$}
            \State $h^t_p \gets RNN_p(h^t_p,[h^t_q,h^t_l])$
            \State $\widetilde{m}^{t+1}_{p,q} \gets h^t_p $
            \State $\widetilde{m}^{t+1}_{p,l} \gets h^t_p $
        \EndFor
        \State $h^{t+1}_p \gets h^t_p $
    \EndFor
    \ForEach {$q \in Q$}
        \State $\widetilde{m}^{t+1}_{q} \gets \sum_{p:k \in k}  \widetilde{m}^{t+1}_{p,k}$
        \State $h^{t+1}_q \gets U_q(h^t_q,\widetilde{m}^{t+1}_{q})$
    \EndFor
    \ForEach {$l \in L$}
        \State $\widetilde{m}^{t+1}_{l}  \gets RNN_l(h^t_l,h^{t+1}_q \forall q \in j(l))$
	 \State $h^{t+1}_l \gets \widetilde{m}^{t+1}_{l}$
    \EndFor
\EndFor
\State $\hat{y_p} \gets F_p(h_p)$
\end{algorithmic}
\label{alg:architecure}
\end{algorithm}

In Algorithm \ref{alg:architecure}, the loop from line 6, and lines 12 and 15 represent the different message-passing operations that exchange the information encoded in the hidden states between queues, links and paths. Likewise, lines 10, 13 and 16 are update functions that incorporate the newly collected information into the hidden states. 

This architecture provides flexibility to represent any routing scheme and queue policy configuration. This is achieved by the direct mapping of $P$ to specific message passing operations between queues, links and paths. Thus, each path collects messages from all the queues and links included in it (loop from line 5), then each queue receives messages from all the paths containing it (line 11) and similarly, each link collects information from all the queues that inject traffic in it (line 14). Given that the order of paths that traverse a queue does not matter, a summation is used to aggregate the paths' hidden states on queues. However, in the case of links and queues, there is a sequential dependence. For this reason, we use a Recurrent Neural Network (RNN) to aggregate the sequences of queues and links on the paths' hidden states. Similarly, the model aggregates the queue states on their related links using an RNN, as the order of queues is important to understand and model the queue policy (e.g., the priority order).

Finally, in Algorithm \ref{alg:architecure}, the function $F_p$ (line 17) represents a readout function, which predicts the mean per-path delay ($\hat{y_p}$) using as input the paths' hidden states $h_p$. Particularly, we modeled the readout function $F_p$ with a fully-connected neural network using SELU activation functions.

\section{Evaluation of the accuracy of the Model}
In this section, we analyze the accuracy of the proposed GNN model to estimate the per-source-destination delay in a wide variety of topologies, routing schemes and traffic intensities. 

\subsection{Simulation Setup}\label{sim_setup}
%- Copy from RouteNet JSAC (https://ieeexplore-ieee-org.recursos.biblioteca.upc.edu/abstract/document/9109574) $\rightarrow$ Adapt capacities, bitrates...
To train our model we built a ground truth with a packet-level network simulator (OMNeT++ v5.5.1~\cite{varga2001discrete}). For each simulation, the traffic, queue policies, and routing scheme is chosen randomly according to the ranges defined below. Then the mean end-to-end delay for every source-destination pair is measured.

\subsubsection{Traffic}
Our traffic model follows a similar approach as in~\cite{rusek2019unveiling}. Particularly, we generate input traffic matrices ($TM$) as follows:
\begin{equation}
\label{eq:tm}
TM(S_i,D_j) = \frac{U(0.1,1) \times TI}{N-1}  \quad   \forall i,j \in nodes, i \neq j 
\end{equation}

Where U(0.1,1) is a uniform distribution in the range $[0.1, 1]$, $N$ is the number of nodes in the network topology and $TI$ is a tunable parameter that indicates the overall traffic intensity in the simulation. $TI$ represents how congested is the network, in our dataset it ranges from 400 to 2000 bits per time unit. Being 400 the lowest congested network (with 0\% packet loss) and 2000 a highly congested network with $\approx$ 3\% of packet loss. The inter-packet arrival time is modelled by a Poisson process, with the mean derived from $TM$. The packet size follows a bimodal distribution commonly used in other works~\cite{sinha2007internet}. Finally, we assign randomly a Type of Service (ToS) label to each source-destination traffic flow \mbox{(ToS$\in$[0-9])}.

 %In each source-destination pair, inter-packet arrival times are modelled with an exponential distribution whose mean is derived from the traffic defined in $TM$. Also, packet sizes follow a binomial distribution, where $50\%$ of the packets have a size of 300 bits and the rest of them contain 1700 bits. Additionally, for each source-destination pair, a Type of Service (ToS) between 0 to 9 is selected.

\subsubsection{Queues}

Each output port of nodes is configured with a specific queing configuration. For this, we select randomly on each port one possible option of the following parameters: $(i)$ a queue scheduling policy, that can be First In First Out (FIFO), Strict Priority (SP), Weighted Fair Queueing (WFQ) or Deficit Round Robin (DRR), $(ii)$ a random number of queues 2-5 (except FIFO, which implements always 1 queue) and $(iii)$ random queue size (16, 32 or 64 packets). For WFQ and DRR, we also define a set of random queue weights that add up to 1. Finally, we also map randomly ToS classes to specific queues on each port.

%The built ground truth has its focus on adding queue scheduling policies, where each node is configured with a different scheduling policy and a different number of priority queues. Particularly, the following policies were used: First In First Out (FIFO), Strict Priority (SP), where packets in queues with more priority are transmitted first. Weighted Fair Queueing (WFQ) and Deficit Round Robin (DRR).

%Each node is configured randomly with one of the queue scheduling policies (FIFO, SP, WFQ or DRR) and assign randomly some queues between 2 to 5 except for FIFO that only have one queue. For WFQ and DRR, we also define a set with 5 queue weights profiles for each possible configuration based on the number of queues. The mapping between ToS and queues is also randomly selected between a set of 100 possible configurations. Furthermore, the size of the queues of a node is randomly selected between 16, 32 or 64 packets per queue. A total of 2000 scheduling configurations (policies, weight profiles, number of queues, queue size) are generated with this criterion for each topology, and each simulation selects randomly one of these configurations.

\subsubsection{Topologies}
In order to train and evaluate the model, we use 3 different real-world topologies. The first one, a 14-node and 21-link (NSF network topology \cite{hei2004wavelength}), The second one, a 24-node and 37 links (GEANT topology \cite{barreto2012fast}). And the third one, a 17-node and 26 links called German Backbone Network (GBN)~\cite{pedro2011performance}.

\subsection{Training and Evaluation}
%- Describe all the datasets (training, test, n. samples, combinations of traffic matrices + routings + queue scheduling configurations)

%- Cost statistics: Add time to train the model (over the training dataset) and describe the setup (e.g., GPU NVIDIA XXX, CPU...). And the time to make an inference on a network with 100 nodes?

We implement the GNN model using TensorFlow. The source code and all the training/evaluation datasets used in this paper will be publicly available upon acceptance. To train the model, we used NSFNET and GEANT topologies as described before. We choose GBN for evaluation. For each topology, 100,000 random samples are used, which gave us a total of 200,000 samples for training and 100,000 samples for evaluation. 

Our model has two relevant hyper-parameters that can be fine-tuned: $(i)$ The size of the hidden states $h_q$, $h_l$ and $h_p$ and $(ii)$ the number of message passing iterations ($T$). Based on early experiments we found in our case a size of 32 for the three different types of hidden states and $T=8$ iterations leads to good accuracy.

\begin{figure}[t]
\centerline{\includegraphics[width=\columnwidth]{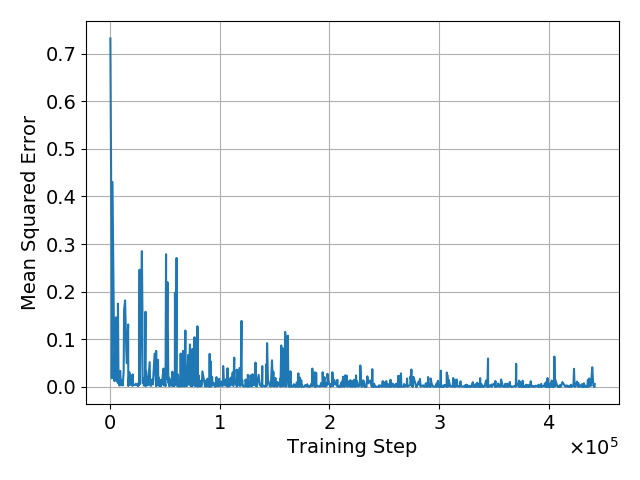}}
\caption{Training loss obtained during the training phase.}
\label{fig:training_loss}
\end{figure}

We choose the Mean Squared Error (MSE) as a loss function, which is minimized using an Adam optimizer with an initial learning rate of 0.001 and a decay rate of 0.6 executed every 80,000 steps. In addition to this, we added an $L2$ regularization loss of $\lambda{=}0.1$. Figure \ref{fig:training_loss} shows how the training loss decays as the training evolves. The training was executed in a testbed with a GPU Nvidia GeForce GTX 1080 Ti for {$\approx$}2 epochs (i.e., {$\approx$}450,000 training steps). In total, the training phase took around 12 hours ({$\approx$}9.25 samples per second).

\begin{table}[]
\centering
\caption{Summary of the evaluation accuracy.}
\label{tab:eval_results}
\resizebox{0.6\columnwidth}{!}{%
\begin{tabular}{@{}c|ccc@{}}
\toprule
      & GBN    & NSFNET & GEANT  \\ \midrule
$MRE$   & 3.88\% & 2.59\% & 3.01\% \\
$R^2$ & 0.81   & 0.99   & 0.95   \\ \bottomrule
\end{tabular}%
}
\end{table}

Table \ref{tab:eval_results} summarizes the evaluation results in the three topologies described in Section III A. Note that the model was only trained with samples from GEANT and NSFNET, while GBN was not seen during the training.  As the proposed GNN can be analyzed as a regression model, we provide two evaluation statistics: (i) the Mean Relative Error (MRE) and (ii) the percentage of variance explained by the model ($R^2$). As we can observe, the model shows good accuracy even in GBN \mbox{-- the} network not seen during training ($MRE=3.88\%$). This high accuracy reveals the capability of the proposed GNN architecture to generalize over different topologies never seen during the training phase.

\section{Related Work}
%- See Related work JSAC paper (https://ieeexplore-ieee-org.recursos.biblioteca.upc.edu/abstract/document/9109574)
%- Four different types of solutions: 1) Fluid models (very simple), 2) analytical models (queuing theory, and network calculus) $\rightarrow$ lightweight but inaccurate, 3) discrete-event network simulation (e.g., ns-3, OMNet) $\rightarrow$ accurate but extremely costly, 4) Recent attempts with NN: traditional NN architectures ("Understanding the modelling of computer network delays using neural networks". "deep-Q"), and with GNN (Krzysztof MPNN, Geyer, RouteNet)

One of the fundamental goals of network modelling is to provide a cost function that is then used for optimization. Over the years, many attempts have been made to obtain good cost functions. This includes fluid models, analytical models (e.g., queuing theory, network calculus) and discrete-event network simulators (e.g., ns-3, OMNet++). 

Among all the existing techniques, queuing theory is possibly the most popular. As an example, \cite{pioro2004routing} obtains an objective function based on the linearization of well-known queuing theory results. Alternatively, fluid models are efficient and popular for some congestion control problems. However, they make important simplification assumptions and, as shown before, it leads to considerable innacuracies~\cite{eun2007limitation}. Likewise, network calculus operates over the worst-case scenarios of networks. Thus, these types of scenarios are rarely observed in operational environments. As a result, this kind of techniques can often lead to poor performance compared to the use of accurate models as the one proposed in this paper.

The use of Deep Learning for network modelling is a topic recently addressed by the research community~\cite{wang2017machine,xiao2018deep,mestres2018understanding}. Existing proposals mainly use common artificial neural networks (e.g., feed-forward NNs, convolutional NNs). The main limitation of these works is that they do not generalize to other topologies and configurations (e.g., routing).

In the context of GNN-based network modelling, RouteNet \cite{rusek2019unveiling,meng2020interpreting} is arguably the closest model to our work. However, this model does not support different queuing policies. Thus, it produces inaccurate delay estimates in scenarios with complex queue scheduling configurations. 

\section{Conclusions}

In this paper, we have proposed a new Graph-based Deep Learning architecture for network modeling. The main novelty of this model is that it is able to predict the impact of arbitrary queuing policies on network performance, while generalizing successfully to unseen network scenarios. Although in this paper we evaluate the model to predict the per-path delay, it can be easily extended to support other QoS metrics such as jitter or packet loss. More importantly, the proposed model is fast compared to other accurate network modeling tools like packet-level simulators.

\bibliographystyle{unsrt}
\bibliography{sample}

\end{document}